# Constructor Theory

David Deutsch

Centre for Quantum Computation, The Clarendon Laboratory,
University of Oxford
and
Future of Humanity Institute,
University of Oxford

Constructor theory seeks to express all fundamental scientific theories in terms of a dichotomy between *possible* and *impossible* physical transformations – those that can be caused to happen and those that cannot. This is a departure from the prevailing conception of fundamental physics which is to predict what *will happen* from initial conditions and laws of motion. Several converging motivations for expecting constructor theory to be a fundamental branch of physics are discussed. Some principles of the theory are suggested and its potential for solving various problems and achieving various unifications is explored. These include providing a theory of information underlying classical and quantum information; generalising the theory of computation to include all physical transformations; unifying formal statements of conservation laws with the stronger operational ones (such as the ruling-out of perpetual motion machines); expressing the principles of testability and of the computability of nature (currently deemed methodological and metaphysical respectively) as laws of physics; allowing exact statements of emergent laws (such as the second law of thermodynamics); and expressing certain apparently anthropocentric attributes such as knowledge in physical terms.

## 1. Introduction

*1.1 Construction tasks*

Consider an automated factory for producing goods to a particular specification. Though its purpose may only be to *produce* those goods, the laws of physics imply that it must *transform* something into them and, typically, also use other resources and produce waste products. Very few such transformations happen spontaneously; that is to say, almost all require a *constructor*, which I shall define as anything that can cause transformations in physical systems without undergoing any net change in its ability to do so. I shall call those physical systems the constructor's *substrates*:

$$\text{input state of substrate(s)} \xrightarrow{\text{constructor}} \text{output state of substrate(s)}. \qquad (1)$$

A transformation, regarded as being caused by a constructor, I call a *construction.*

Constructors appear under various names in physics and other fields. For instance, in thermodynamics, a heat engine is a constructor because of the condition that it be capable of 'operating in a cycle'. But they do not currently appear in *laws* of physics. Indeed, there is no possible role for them in what I shall call the *prevailing conception* of fundamental physics, which is roughly as follows: everything physical is composed of elementary constituents such as particles, fields and spacetime; there is an initial state of those constituents; and laws of motion determine how the state evolves continuously thereafter. In contrast, a construction (1) is characterised only by its inputs and outputs, and involves subsystems (the constructor and the substrate), playing different roles, and most constructors are themselves composite objects. So, in the prevailing conception, no law of physics could possibly mention them: the whole continuous process of interaction between constructor and substrate is already determined by the universal laws governing their constituents.

However, the constructor theory that I shall propose in this paper is not primarily the theory of constructions or constructors, as the prevailing conception would require it to be. It is the theory of which *transformations*

$$\text{input state of substrates} \;\rightarrow\; \text{output state of substrates} \qquad (2)$$

*can be caused* and which cannot, and why. As I shall explain, the idea is that the fundamental questions of physics can all be expressed in terms of those issues, and that the answers do not depend on what the constructor is, so it can be abstracted away, leaving transformations (2) as the basic subject matter of the theory. I shall argue that we should expect such a theory to constitute a fundamental branch of physics with new, universal laws, and to provide a powerful new language for expressing other theories. I shall guess what some of those laws may be, and explore the theory's potential for solving various problems and achieving various unifications between disparate branches of physics and beyond, and propose a notation that may be useful in developing it.

Causation is widely regarded by philosophers as being at best a useful fiction having no possible role in fundamental science. Hume (1739) argued that we cannot observe causation and therefore can never have evidence of its existence. But here I shall, with Popper (1959, 1963), regard scientific theories as conjectured explanations, not as inferences from evidence, and observation not as a means of validating them, but only of testing them. So Hume's argument does not apply. Nor does the argument (e.g. by Russell 1913) that the fundamental laws of physics make no reference to causes – for that is merely an attribute of a particular way of formulating those laws (namely, the prevailing conception) not of the laws themselves. Moreover, the prevailing conception itself is not consistent about that issue, for the idea of a universal law is part of it too, and the empirical content of such a law is in what it *forbids* by way of testable outcomes (Popper 1959, §31 & §35) – in other words in what transformations it denies *can be caused* to happen, including to measuring instruments in any possible laboratories. Explanatory theories with such counter-factual implications are more fundamental than predictions of what *will* happen. For example, consider the difference between saying that a purported perpetual motion machine *cannot be made* to work as claimed 'because that would violate a conservation law' and that it *won't* work 'because that axle exerts too small a torque on the wheel'. Both explanations are true, but the former rules out much more, and an inventor who understood only the latter might waste much more time trying to cause the transformation in question by modifying the machine.

I provisionally define a *construction task* (or 'task', for short) as a *set* of pairs such as (2), each designating a *legitimate input state* for the task and associating that with a *legitimate output state* for that input. (So constructor theory might be more accurately called *construction task theory*, but I think the shorter name is preferable.) A constructor is *capable of performing* a task $\mathfrak{A}$ if, whenever it is presented with substrates in a legitimate input state of $\mathfrak{A}$, it transforms them to one of the output states that $\mathfrak{A}$ associates with that input. For example, if *p, q, r* and *s* are states of some substrate, then to be capable of performing the task $\{p \to q, q \to r, q \to s\}$, a constructor must invariably produce *q* if presented with *p*, and may produce either *r* or *s* if presented with *q*, and may do anything at all if presented with any other state. A constructor's behaviour when presented with anything other than a legitimate input state is undefined. The *transpose* $\mathfrak{A}^{\sim}$ of $\mathfrak{A}$ is the task with all the input and output states of $\mathfrak{A}$ swapped, so $\{p \to q, q \to r, q \to s\}^{\sim} \equiv \{q \to p, r \to q, s \to q\}$.

Presumably no perfect constructors can exist in nature. A factory is only an approximation to one, as are some of its constituents such as robots and conveyor belts, because of their non-zero error rates (producing something other than the specified output), and because in the absence of maintenance they gradually deteriorate. A task $\mathfrak{A}$ is *possible* (which I write as $\mathfrak{A}^{\checkmark}$) if the laws of nature impose no limit, short of perfection, on how accurately $\mathfrak{A}$ could be performed, nor on how well things that are capable of approximately performing it could retain their ability to do so. Otherwise $\mathfrak{A}$ is *impossible* (which I write as $\mathfrak{A}^{\times}$).

It may be that construction tasks are the primitive entities in terms of which the laws of nature are expressed. In that case, a 'set of ordered pairs of states' would be only a provisional way of conceiving of tasks: ultimately substrates, states and transformations would be understood in terms of tasks, not vice versa.

A construction task specifies only intrinsic attributes of its substrates, not extrinsic ones such as the name or location of a particular instance. So, converting the Eiffel Tower into a billion toy models of itself is not a construction task, but converting a structure with that specification into such models, plus waste products, using a specified source of energy, is.

The specification of a task includes all its necessary inputs and unavoidable outputs, hence the need to specify such things as what energy sources and waste products are permitted. Thus, a constructor is characterised by the effect it would have if it and its substrates jointly constituted a closed system. For instance, a device connected to an external power supply should not qualify as a perpetual motion machine.

The definition of a constructor for a task $\mathfrak{A}$ requires it to be a constructor for $\mathfrak{A}$ again after performing an instance of $\mathfrak{A}$. Its other attributes may change, but what the user must do in order to cause it to perform the task must remain the same. (The term 'user' here is not intended to have any anthropomorphic connotation; the user is whatever presents the constructor with its substrates, causing it to perform its task.) If a machine stops being capable of performing a task $\mathfrak{A}$ after its $N$'th run because its battery has run down, then it is not a constructor for $\mathfrak{A}$ after the $(N-1)$'th run either, because the $N$'th run does not end with it being such a constructor – and so it follows by induction that it never was one. However, the same machine excluding the battery could still be a constructor for a related task whose substrates include a battery whose legitimate input states specify enough charge to perform $\mathfrak{A}$ at least once.

*1.2 Composition of tasks*

A candidate formalism for constructor theory is the *algebra of tasks.* The *parallel composition* $\mathfrak{A} \otimes \mathfrak{B}$ of tasks $\mathfrak{A}$ and $\mathfrak{B}$ is the task of achieving the effect of performing $\mathfrak{A}$ on one substrate and $\mathfrak{B}$ on another. But performing $\mathfrak{A} \otimes \mathfrak{B}$ need not involve actually performing $\mathfrak{A}$ or $\mathfrak{B}$. Indeed, $\mathfrak{A} \otimes \mathfrak{B}$ may be possible even when neither $\mathfrak{A}$ nor $\mathfrak{B}$ is – for instance, when they violate a conservation law by equal and opposite amounts. The same holds for the *serial composition* $\mathfrak{A}\mathfrak{B}$, which is defined such that in cases where every legitimate output state of $\mathfrak{B}$ is a legitimate input state of $\mathfrak{A}$, $\mathfrak{A}\mathfrak{B}$ is the task of achieving the net effect of performing $\mathfrak{B}$ and then immediately performing $\mathfrak{A}$ on the same substrate. (Other cases will not concern us in this paper.) We can also define tasks as networks of sub-tasks, with the outputs of some connected to the inputs of others. A *regular network* is one in which the legitimate outputs of the task at the beginning of each link are the legitimate inputs of the task at its end. This formalism makes it natural to conjecture – and here we come to our first constructor-theoretic law of physics, which I shall call the *composition principle* – that *every regular network of possible tasks is a possible task*.

However, networks of *possible* tasks are not the only subject-matter of constructor theory. A possible task can be composed of impossible ones, as in the case of conservation laws that I have just mentioned. In this respect constructor theory differs from the theory of computation, which it generalises (Section 2.7 below), and from other extensions of the idea of computation, such as modal logics. Those all deal with how to compose *constructions*, which are processes in which constructors perform possible tasks.

Nor is constructor theory intended to implement a positivist or operationalist program of excluding from science everything but statements of which tasks are possible or impossible. On the contrary, even a complete set of such statements, if uninterpreted by explanation, would be unintelligible as a theory about nature. That is why I characterised constructor theory above as the theory of which transformations can or cannot be caused *and why*. Science in the constructor-theoretic conception remains primarily explanatory, as all science should be (Deutsch 2011, ch. 1). Far from excluding anything, I expect constructor theory to permit a wider class of explanations than the prevailing conception (as, for instance, in Section 2.2 below).

*1.3 Constructor theory would underlie all other scientific theories*
The theory of relativity is the theory of the arena (spacetime) in which all physical processes take place. Thus, by its explanatory structure, it claims to *underlie* all other scientific theories, known and unknown, in that requires them to be expressible in terms of tensor fields on spacetime, and constrains what they can say about the motion of those fields. For example, any theory postulating a new particle that was unaffected by gravity (i.e. by the curvature of spacetime) would contradict the general theory of relativity. Another theory that inherently claims to underlie all others is quantum theory, which requires all observable quantities to be expressible in terms of quantum-mechanical operators obeying certain commutation laws. And so, for example, no theory claiming that some physical variable and its time derivative are simultaneously measurable with arbitrary accuracy can be consistent with quantum theory. Constructor theory would, in this sense, underlie all other theories including relativity and quantum theory. The logic of the relationship would be as follows: Other theories specify what substrates and tasks exist, and provide the multiplication tables for serial and parallel composition of tasks, and state that some of the tasks are impossible, and explain why. Constructor theory provides a unifying formalism in which other theories can do this, and its principles constrain their laws, and in particular, require certain types of task to be possible. I shall call all scientific theories other than constructor theory *subsidiary theories*.

In this paper I guess some principles of constructor theory. Some of these guesses (such as the composition principle in Section 1.2 above and the interoperability principle in Section 2.6 below) are natural given the formalism I am proposing, which is itself natural if constructor theory is regarded as a generalisation of the theory of computation (Section 2.7). Others are principles that are already known but not widely regarded as laws of physics, such as the principle of testability (Section 2.11) and what I have called the Turing principle (Section 2.8). If the thesis of this paper is true, then others may emerge from the explicit integration of subsidiary theories, especially quantum theory and relativity, and by guessing improvements to falsified guesses (Section 2.10).

## 2. Motivations

### 2.1 Catalysis

A *catalyst* is a substance that increases the rate of a chemical reaction without undergoing any net chemical change itself. Chemical equations describing catalysis are written like this:

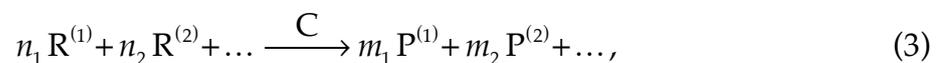

$$n_1 R^{(1)} + n_2 R^{(2)} + \ldots \xrightarrow{C} m_1 P^{(1)} + m_2 P^{(2)} + \ldots, \tag{3}$$

which conforms to the pattern (1) with the catalyst C as the constructor.

Since a catalyst changes only the rate of a reaction, not the position of equilibrium, it is sometimes deemed a mistake to regard catalysts as *causing* reactions. However, that argument would deny that anything causes anything. Even without a factory, the components of a car do spontaneously assemble themselves at a very low rate, due to Brownian motion, but this happens along with countless other competing processes, some of them (such as rusting away) much faster than that self-assembly, and all of them much slower than the assembly effected by the factory. Hence a car is overwhelmingly unlikely to appear unless a suitable constructor is present. So if causation is meaningful at all, catalysts and other constructors do indeed cause their characteristic constructions.

When one is not specifically discussing the catalyst, one usually omits it, describing the reaction as a construction task instead:

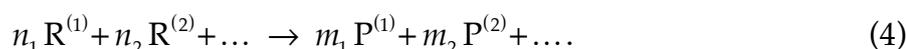

$$n_1 R^{(1)} + n_2 R^{(2)} + \ldots \;\rightarrow\; m_1 P^{(1)} + m_2 P^{(2)} + \ldots. \tag{4}$$

This is convenient because most laws of chemistry are only about the reagents; that is to say, they hold regardless of what the catalyst may be, and hence assert nothing about the catalyst. For example, the *law of definite proportions* requires the coefficients $n_1, n_2 \ldots$ and $m_1, m_2 \ldots$ in (3) or (4) to be integers, depending only on the chemical identities of the reagents and products. It says that *any* catalyst capable of catalysing (4) can do so only for integer values of the coefficients. Similarly, (4) has to balance (expressing the fact that chemical processes cannot create or destroy atoms); it has to scale (be the same whether the terms refer to molecules, moles or any other measure proportional to those); the free energy of the products must not exceed that of the reagents; and so on. All these laws hold whatever causes the reaction while remaining unchanged in its ability to do so.

Imposing the prevailing conception of fundamental physics on chemistry would entail treating the catalyst as another reagent. One would rewrite (3) as

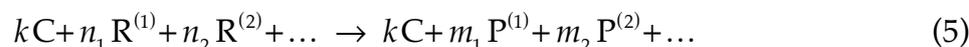

$$k C + n_1 R^{(1)} + n_2 R^{(2)} + \ldots \;\rightarrow\; k C + m_1 P^{(1)} + m_2 P^{(2)} + \ldots \tag{5}$$

for some $k$. But then the catalyst violates the law of definite proportions: since each catalyst molecule may be re-used, (5) can proceed for a huge range of values of $k$. Nor does (5) scale: the minimum number of catalyst molecules for which it outpaces competing reactions is some $k_0$, but for $x$ times the number of reagent molecules, the minimum number may be much lower than $xk_0$, and will depend on non-chemical factors such as the size of the container, again contrary to the law of definite proportions.

The customary distinction between catalysts and other reagents therefore correctly reflects the fact that they are treated differently by laws of nature – in this case, laws of chemistry. But there is no significant distinction between catalysts and other *constructors*. For example, the synthesis of ammonia, $3\text{H}_2 + 2\text{N}_2 \xrightarrow{\text{Fe}} 2\text{NH}_3$, will not happen in empty space, because at near-zero pressure the process of diffusing away is much faster than the chemical reaction. Hence a container or equivalent constructor is among the conditions required in addition to the catalyst. Indeed, some catalysts work by *being* microscopic containers for the reagents.

Chemical catalysis has natural generalisations. Carbon nuclei are catalysts for nuclear reactions in stars. A living organism is both a constructor and a product of the construction that is its life-cycle which, for single-celled photosynthesising organisms, is simply:

$$\text{small molecules} + \text{light} \xrightarrow{\text{cell}} \text{cell} + \text{waste products}. \qquad (6)$$

Inside cells, proteins are manufactured by ribosomes, which are constructors consisting of several large molecules. They function with the help of smaller catalysts (enzymes) and water, using ATP as fuel:

$$\text{amino acids} + \text{ATP} \xrightarrow{\text{RNA} + \text{ribosome} + \text{enzymes} + \text{H}_2\text{O}} \text{protein} + \text{AMP} + \text{waste products}. \quad (7)$$

I mention this reaction in particular because the RNA plays a different role from the other catalysts. It specifies, in a code, which protein shall be the product on a given occasion. Thus, the catalysts excluding the RNA constitute a *programmable* constructor. The general pattern is:

$$\text{input state of substrates} \xrightarrow[\text{programmable constructor}]{\overset{\text{program}}{\Downarrow}} \text{output state of substrates}. \quad (8)$$

Constructor theory is the ultimate generalisation of the idea of catalysis.

*2.2 Constructor-independent laws of physics*

In physics, too, most laws are about substrates only: they hold for all possible constructors. But the prevailing conception disguises that. For example, one way of formulating a conservation law in the prevailing conception is that for every isolated physical system **S**, a certain quantity $Q(\mathbf{S})$ never changes. But since an isolated system never deviates from a particular trajectory, almost all its attributes (namely, all deviations from that trajectory) remain unchanged, not just conserved quantities. For instance, the total energy of the atoms of an isolated crystal remains unchanged, but so does their arrangement, and only the former invariance is due to a conservation law. In constructor-theoretic terms, the arrangement could be changed by a constructor while the energy could not.

More generally, what makes the attributes that we call 'conserved quantities' significant is not that they cannot change when a system is isolated, but that they cannot *be* changed, *whether or not* the system is isolated, without depleting some external resource. In the prevailing conception, one addresses this by considering additive conserved quantities (ones for which $Q(\mathbf{S}_1 \oplus \mathbf{S}_2) = Q(\mathbf{S}_1) + Q(\mathbf{S}_2)$, where $\mathbf{S}_1 \oplus \mathbf{S}_2$ denotes the composite system of $\mathbf{S}_1$ and $\mathbf{S}_2$). Then if $\mathbf{S}_1 \oplus \mathbf{S}_2$ is isolated, the conservation law says that regardless of how $\mathbf{S}_1$ and $\mathbf{S}_2$ may interact, $Q(\mathbf{S}_2)$ cannot change unless $Q(\mathbf{S}_1)$ changes by an equal and opposite amount. But that is not enough to imply the most interesting property of conserved quantities, namely that this balancing act cannot continue indefinitely. If it could, then $\mathbf{S}_1$ could be used to provide an everlasting supply of $Q$ to successive instances of $\mathbf{S}_2$, while $Q(\mathbf{S}_1)$ took ever more negative values. If $Q$ was energy, $\mathbf{S}_1$ would then constitute a perpetual motion machine of the first kind. So, in the prevailing conception, the principle of the conservation of energy alone does not rule out perpetual motion machines. One needs an additional fact: energy is bounded below. Moreover, since it is unbounded above, this still does not rule out a machine that would sequester energy indefinitely from successive substrates. That cannot exist for another reason: to increase the energy of any system $\mathbf{S}_1$ without limit, the actions that the user needs to perform keep changing, thus disqualifying $\mathbf{S}_1$ from being a constructor. For instance, even if $\mathbf{S}_1$ incorporated the most compact possible means of sequestering energy, namely a black hole, its size and gravitational effects would increase whenever it was used, requiring increasing resources to prevent it from impairing its users' functionality.

Similar considerations apply to electric charge (which is neither bounded below nor above) and to all other conserved quantities: in the prevailing conception, conservation laws do not rule out that a finite substrate could be an unlimited source or sink of the quantity, yet in no case can it. Thus, for each conserved quantity $Q$ the prevailing conception conceals a deeper regularity – a stronger conservation law – which can be succinctly expressed in constructor theory and holds whether $Q$ is additive or not: *every task that would change the (overall) Q of its substrates is impossible*.

*2.3 Irreversibility*

All the laws of thermodynamics have informal constructor-theoretic versions (i.e. versions saying that certain tasks are impossible), but for the second law, *only* informal or approximate versions currently exist. That is because of its notorious *prima facie* conflict with the reversibility of all known laws of motion, and hence with the prevailing conception. All existing proposals for resolving this depend, explicitly or implicitly[1], on ensemble-averaging or coarse-graining to define entropy and heat, which makes the connection with physical states vague and arbitrary. A similar problem exists for equilibrium, referred to by the zeroth law. Expressing those laws in terms of tasks may make all the difference, because in the prevailing conception, the reversibility of a composite process follows logically from that of its constituent processes, and hence we cannot expect the second law to be an exact law of nature at all. But in constructor theory there is no reason why a possible task, such as converting work entirely into heat, should not have an impossible transpose. So it seems reasonable to hope that the informal statements of all the laws of thermodynamics, as well as concepts analogous to heat, entropy and equilibrium, could be made precise within constructor theory by formalising existing informal

---

[1] For example, the most formal treatments of the second law, such as Carathéodory's, simply assume the existence of 'macro'-states.

statements such as 'it is impossible to build a perpetual motion machine of the second kind'.

Similarly, quantum decoherence is not a property of the exact quantum state of the universe but only of coarse-grained approximations. Yet it is fundamental to the structure of the multiverse (Deutsch 2002) and is related to thermodynamics in imperfectly understood ways. It, too, may eventually be understood as an exact attribute of tasks – for example as a statement of the impossibility of tasks that would transmit information (Section 2.6 below) between universes.

*2.4 Von Neumann's approach*

Before the discovery of the structure of DNA, von Neumann (1948) wondered how organisms can possibly reproduce themselves faithfully and evolve complex adaptations for doing so. He realised that an organism must be a *programmable* constructor operating in two stages, namely copying its program and executing it to build another instance. He tried to model this using simplified laws of physics – thus founding the field of cellular automata – but without success: it was too complicated. He also introduced an important constructor-theoretic idea, namely that of a universal constructor (3.11 below), but he made no further progress in constructor theory because, by retreating to cellular automata, he had locked himself into the prevailing conception and also abstracted away all connections between his theory and physics.

*2.5 What is the initial state?*

The prevailing conception regards the initial state of the physical world as a fundamental part of its constitution, and we therefore hope and expect that state to be specified by some fundamental, elegant law of physics. But at present there are no exact theories of what the initial state was. Thermodynamics suggests that it was a 'zero-entropy state', but as I said, we have no exact theory of what that means. Cosmology suggests that it was homogeneous and isotropic, but whether the observed inhomogeneities (such as galaxies) could have evolved from quantum fluctuations in a homogeneous initial state is controversial.

In the constructor-theoretic conception, the initial state is not fundamental. It is an emergent consequence of the fundamental truths that laws of physics specify, namely which tasks are or are not possible. For example, given a set of laws of motion, what exactly is implied about the initial state by the practical feasibility of building (good approximations to) a universal computer several billion years later may be inelegant and intractably complex to state explicitly, yet may follow logically from elegant constructor-theoretic laws about information and computation (see Sections 2.6 and 2.8 below).

The intuitive appeal of the prevailing conception may be nothing more than a legacy from an earlier era of philosophy: First, the idea that the *initial* state is fundamental corresponds to the ancient idea of divine creation happening at the beginning of time. And second, the idea that the initial state might be a logical consequence of anything deeper raises a spectre of teleological explanation, which is anathema because it resembles explanation through divine intentions. But neither of those (somewhat contradictory) considerations could be a substantive objection to a fruitful constructor theory, if one could be developed.

*2.6 Information*

Since all known laws of motion are logically reversible (the initial state of any isolated system is entailed by its final state, as well as vice versa), an irreversible computational task on a computer's memory **M** is a reversible one on some $\mathbf{M} \oplus \mathbf{W}$, where **W** is an additional substrate that carries away waste information. So for present purposes we can confine attention to *reversible computers*, namely constructors some of whose substrates are interpreted as *reversible information-processing media*. Such a medium can be defined in purely constructor-theoretic terms: it is a substrate with a set *S* of at least two states such that for all permutations (i.e. reversible computations) $\Pi$ on *S*,

$$\left( \bigcup_{x \in S} \{x \to \Pi(x)\} \right)^{\checkmark}. \tag{9}$$

Implicitly, defining such media defines information itself[1].

Theories of information and computation exist in their own right, independently of theories of particular physical systems, because there exist substrate-independent truths about the physical world – which, again, have their natural expression in constructor theory. The *interoperability principle for information* is one: If $\mathbf{S}_1$ and $\mathbf{S}_2$ are reversible information-carrying media satisfying (9) with $S = S_1$ and $S = S_2$ respectively, then $\mathbf{S}_1 \oplus \mathbf{S}_2$ is a reversible information-carrying medium with $S$ as the set of all ordered pairs $\{(p,q) | p \in S_1, q \in S_2\}$ (see 3.9 below). In particular, this implies that for any medium, if its states $S$ satisfying (9) are used to store information, copying that information to (a blank instance of) any other medium of the same or greater capacity is a possible task.

*2.7 Generalising computation*

Other construction tasks involving information are *measurement* (in which a physical system is an input and the desired output is information about the system) and *preparation* (in which the desired output is a physical system meeting a criterion specified in the input). So we have the following classification:

|  |  | Output |  |
|---|---|---|---|
|  |  | **Abstract** | **Physical** |
| **Input** | **Abstract** | Computation | Preparation |
|  | **Physical** | Measurement | Other construction |

Table 1: A classification of constructions

---

[1] In this paper the term 'information' means classical information. A more general notion, which includes both quantum information and physically irreversible media, can likewise be defined in purely constructor-theoretic terms. This will be discussed in a future paper.

While three of the four phenomena in Table 1 have been extensively studied, the lower-right one has received scant attention. It is therefore natural to try to unify all four under a single theory: constructor theory.

*2.8 The computability of nature*

The theory of computation was originally intended only as a mathematical technique for studying proof (Turing 1936), not a branch of physics. Then, as now, there was a widespread assumption – which I shall call the *mathematicians' misconception* – that what the rules of logical inference are, and hence what constitutes a proof, are *a priori* logical issues, independent of the laws of physics. This is analogous to Kant's (1781) misconception that he knew with certainty what the geometry of space is. In fact proof and computation are, like geometry, attributes of the physical world. Different laws of physics would in general make different functions computable and therefore different mathematical assertions provable. (Of course that would make no difference to which mathematical assertions are *true*.) They could also make different physical states and transformations simple – which determines which computational tasks are tractable, and hence which logical truths can serve as rules of inference and which can only be understood as theorems.

Another way of exhibiting the mathematicians' misconception is to pick some formal system that human brains are capable of instantiating to a good approximation (say, the theory of Turing machines), and then to designate certain entities within that system as the only genuine proofs. In that case, if the laws of physics were such that a physical system existed whose motion could be exploited to establish the truth of a Turing-unprovable proposition, the proposition would still not count as having been genuinely proved. Such conceptions of proof are essentialist (since they assign the attribute 'genuine' by fiat), and anthropocentric (since they require us to draw different conclusions from the outcomes of computations according to whether they can be instantiated in human brains). But if the angles of your formal system's 'triangles' always sum to two right angles, it does not follow that those of physical triangles do. And if your system classifies a problem as 'intractable', it may still be easily soluble, even if not by human brains. In those ways and others, Turing's intuitions about computation, like Kant's about geometry, did not match reality.

They are contradicted by the quantum theory of computation and the general theory of relativity respectively, both of which are branches of physics. Even if they had been true, they would still have been claims about the laws of physics, not *a priori* truths.

The fundamental status of general relativity is reflected in the mathematical formalism of physics. For instance, laws of motion are expressed in terms of tensor fields on spacetime. But computability is an even more pervasive regularity in nature: every physical system is described by mathematical functions drawn from the same infinitesimally small class. Yet that regularity is not explicit in any current formulation of the laws of physics. Under the mathematicians' misconception, it was mistaken for a (meta-)mathematical principle, the *Church–Turing conjecture*, namely that the universal Turing machine (an abstract object) is a *universal computer*: one whose *repertoire* (the set of all computations it can be programmed to be capable of performing) includes that of every other computer. But without adducing laws of physics one cannot prove or disprove that conjecture or even state it precisely, because there is no *a priori* (physics-free) way of characterising 'every other computer'.

Furthermore, if some Turing machine can simulate a world **W**, that does not imply that **W**'s laws of physics would be computable by anything inside **W**; yet *that* is the regularity that we have discovered in our world. If 'computability' is to express that regularity, and if its theory is to serve as the foundation of proof theory, it must refer to what physical computers can do, not abstract ones. The principle of the computability of nature must be that a computer capable of simulating any physical system *is physically possible*. (This is a slightly improved version of what I have called the *Turing principle* (Deutsch 1985).) Because computers are complex, composite entities, this principle has no natural expression in the prevailing conception. But it has one in terms of tasks, namely that for programmable computers, *the union of all possible repertoires* (as defined in 3.7 below) *is a possible repertoire*.

*2.9 Why the theory of computation isn't the whole of physics*

There is a sense in which the top-left cell of Table 1 (computation) contains the whole table (the whole of physics), namely: for every possible motion of every object permitted by the laws of physics, there exist virtual-reality renderings in which the simulated object mimics all the observable properties of the real one with arbitrary accuracy short of perfection. Thus any physically possible process corresponds to some set of computer programs; moreover every program, when running, is a physical process. Does the theory of computation therefore coincide with physics, the study of all possible physical objects and their motions?

It does not, because the theory of computation does not specify *which* physical system a particular program renders, and to what accuracy. That requires additional knowledge of the laws of physics. In particular, most programs, if regarded as virtual-reality renderings of physical objects, only render other computers running an equivalent program, and only their computational variables at that. So the theory of computation is only a branch of physics, not vice versa, and constructor theory is the ultimate generalisation of the theory of computation.

*2.10 Laws and principles*

The deepest known laws of nature, sometimes called 'principles', are meta-laws, constraining other laws rather than the behaviour of physical objects directly. For example, the principle of the conservation of energy does not say what forms of energy exist, nor what the energy of any particular system is. Rather, it asserts that for any system **S**, the object-level laws (those governing **S** and its interactions with other systems) define a quantity that has the usual properties of energy.

Principles purport to constrain all true laws, known and unknown. But there is no way of deducing such an implication from laws expressed in the prevailing conception. At most one can add it informally, or prove that existing laws conform to the principle. But all laws of constructor theory are principles; and when they call a task possible, that rules out the existence of insuperable obstacles to performing it, even from unknown laws.

It is sometimes claimed that principles are untestable. An object-level theory is testable if it makes predictions which, if the theory were false, could be contradicted by the outcome of some possible observation, which is then said to *falsify* the theory. Now, mathematics alone determines whether an object-level law *L* obeys a principle *P*. But if it does (so the argument goes), the experimental falsification of *L* would not falsify *P*, because it would not rule out that some unknown law *L′* conforming to *P* might be true. And if *L* violates *P*, the experimental corroboration of *L* would not falsify *P* either, because then some alternative explanation might still satisfy *P*. For example, experiments in the 1920s, interpreted according to the then-prevailing theories (*L*) of what elementary particles exist, implied that the energy of a nucleus before beta decay is greater than the total energy of the decay products. But that did not falsify the principle (*P*) of the conservation of energy: Pauli guessed (*L′*) that energy was being carried away by unknown particles (nowadays called antineutrinos). No experimental results could ever rule out that possibility – and the same holds for any principle.

But this supposed deficiency is shared by all scientific theories: Tests always depend on *background knowledge* – assumptions about other laws and about how measuring instruments work (Popper 1963, ch. 10 §4). Logically, should any theory fail a test, one always has the option of retaining it by denying one of those assumptions. Indeed, this has been used as a critique of the very idea of testability (Putnam 1974). But scientific theories are not merely predictions. They are, primarily, *explanations*: claims about what is there in the physical world and how it behaves. And the negation of an explanation is not an explanation; so a claim such as 'there could be an undetected particle carrying off the energy' is not a scientific theory. Nor is 'perhaps energy is not conserved'. Those are research proposals, not explanations. Consequently the methodology of science includes the rule that any proposal to modify a background-knowledge assumption *must itself be a good explanation*[1]. So, if the only explanatory implication of Pauli's suggestion had been to save the principle of the conservation of energy, both would have been abandoned (as the principle of parity invariance *was* abandoned as a result of other experiments on beta decay). In the event, it was soon used to account for other observations and became indispensable in the understanding of nuclear phenomena, while no good explanation contradicting energy conservation was found.

---

[1] For a discussion of what makes an explanation 'good', see Deutsch (2011) ch. 1.

Quite generally: if there is a phenomenon for which there is a good explanation that violates a principle *P*, and the phenomenon falsifies all known object-level explanations conforming to *P*, then the methodology of science mandates rejecting *P*. Thus *P* is just as exposed to falsification by experiment as any object-level theory.

Another claim that is made in principled opposition to principles is that they are inherently less fundamental than object-level theories because they do not explain what brings about the regularities that they describe (Brown & Timpson 2006). Einstein once believed (1908) that the special theory of relativity, which he had expressed in terms of principles such as that of the constancy of the speed of light, could only be made 'definitive' and 'satisfactory' if it was supplemented by a theory of how interatomic forces make rods contract and clocks slow as the theory predicts. This is an error in its own right – the same error, in four dimensions instead of three, as expecting interatomic forces to be responsible for the apparent smallness of a distant object. But again, object-level theories have the same alleged deficiency: no theory can explain everything it refers to. And, perversely, the more a principle expresses true, fundamental, constraints on as-yet-unknown laws, the less 'definitive' and 'satisfactory' it is in Einstein's 1908 view. In practice, principles are indispensable in the discovery of new laws precisely because they express truths about nature that are missing from object-level theories.

Singling out principles as illegitimate is again reminiscent of the archaic conception of laws of nature as divine commands: it is easier to imagine invisible angels being commanded to enforce local equations than to ensure that the net effect of their efforts will conform to particular principles. Similarly, in the prevailing conception, principles are obeyed only 'because' local laws of motion are. But there is no such chain of command in reality. Events may cause other events, but principles neither cause object-level laws nor vice versa. And while principles can explain why laws are as they are, object-level laws cannot possibly explain principles. So, if you reject principles on principle, then regardless of how well the hopes expressed here may eventually be vindicated, you will regard constructor theory as mere commentary on the laws of nature, its own laws mere coincidence. You may thereby be denying yourself, on principle, the solutions to some longstanding problems at the foundations of physics. It *that* principle self-consistent?

*2.11 The principle of testability*

It has become a standard methodological rule in science (Popper 1959 §10) to consider only testable explanations of physical phenomena. This is equivalent to assuming that the laws of nature do not make their own testing, or each other's, an impossible task. And that is itself a principle of nature: the *principle of testability*. It is as universal as the Turing principle, and at least as alien to the prevailing conception, since experimental tests require complex constructors such as measuring instruments and computers.

And again, testability fits naturally into the constructor-theoretic conception: every experimental test is of some prediction that performing certain transformations on given substrates is an impossible task. For example, the laws of chemistry say that

$$\left( \bigcup_{k \geq 3} \{3\,\mathrm{H}_2 + \mathrm{N}_2 \to k\,\mathrm{NH}_3\} \right)^{\text{✘}}. \tag{10}$$

(That is because, given the definition of 'capable of performing' in Section 1.1, and since every constituent task of the union in (10) has the same input state, any constructor capable of performing $\{3\,\mathrm{H}_2 + \mathrm{N}_2 \to k\,\mathrm{NH}_3\}$ for some particular $k \geq 3$ would necessarily also be capable of performing the whole task in (10).)

The principle of testability, like all the other principles of constructor theory that I suggest in this paper, is itself testable in the same sense as existing principles of physics (2.9 above).

*2.12 The set of all conceivable transformations*

Consider any two physical objects, and imagine one being transformed into the other. Most such conceivable transformations are forbidden by laws of physics, and so the prevailing conception has no role for them. Yet the empirical content of a scientific theory is in what it forbids (and the explanatory content consists of explaining why).

In the prevailing conception, constructions are untypical events; most of what happens in the world is spontaneous, while in the constructor-theoretic conception, it is spontaneous transformations that are untypical. They are the ones that can be 'caused' by a timer alone (3.12 below). Moreover, the simpler the task is (in terms of physical states or transformations), the more complex a constructor for it typically has to be (in terms of physical objects). For instance, the simplest constructor that would produce one Higgs particle starting from hydrogen atoms may well be a particle accelerator kilometres across. Thus, the constructor-theoretic conception regards physically simple things as simple, while the prevailing conception typically regards them as side-effects of very complex ones. In practice, we do usually abstract away that complexity, and when we do, we are departing from the prevailing conception and thinking in constructor-theoretic terms.

*2.13 Counterfactual possibility*

When an engineering study concludes that a proposed structure *would* collapse if it were built, we regard this as meaningful whether the structure is built or not. Indeed, if it is never built, that very prediction may be the reason. 'Counterfactual' predictions of that kind seem to be essential in science and everyday life, and even in the idea of a law of physics, as I mentioned in Section 1.1. Yet what, in reality, do they mean? That is part of a more general *problem of counterfactuals*. Some counterfactual statements are simply under-determined – as, for instance, in Quine's (1960) puzzle that "If Caesar were in charge [in the Korean War]", the statements "he would have used nuclear weapons" and "he would have used catapults" are both arguably true. In such cases, adding enough extra details like 'and had been informed about modern weapons' removes the ambiguity. But the problem with the engineering prediction does not seem to be ambiguity; it is about what the apparently unambiguous commonsense meaning can possibly be asserting about the physical world. For once we know the prediction, and hence know that the structure will not be built, we also know that '*if* it is built, Earth will be attacked by beings from another galaxy'.

I have previously suggested (Deutsch 1997, ch. 11) that this problem is solved in the Everett interpretation of quantum theory, where propositions like 'it would have happened' can be understood as 'it happened in another universe'. However, this only works for events that do happen somewhere in the multiverse (such as a statistical fluctuation bringing into existence a person with specified Caesar-like attributes and causing everyone to regard him as a military commander). It does not work for laws of physics. For example, the principle of testability is doubly counterfactual: it is that a certain task (the test, resulting in a falsifying outcome) must be *possible* if the theory *is false*, but theories expressing actual laws of nature – the very theories to which the principle primarily applies – are not false anywhere in the multiverse.

Computability is another counterfactual attribute: a computable function $f$ is one for which a universal computer *could* be programmed to evaluate $f(i)$ for any integer input $i$, even if no such program ever runs, or if there are integers that will never be its input. Correspondingly, a universal computer is defined in terms of its *ability* to perform a class of tasks, whether it ever does so or not. The traditional theory of computation had no counterfactual import, but only because it had no factual import: its universal computer only 'existed' in the sense of being a certain type of mathematical object. But as I have argued, to regard computation as a physical process, one must give physical meanings to its defining attributes – and some of those are counterfactual.

Counterfactuals are central to constructor theory: not only is every assertion that a task is possible or impossible counterfactual, constructor theory requires subsidiary theories to provide characterisations of *impossible* tasks (Section 3.6 below) and hence of impossible events and processes. Moreover, it seems that all counterfactual assertions that are essential in science or everyday life are expressible in terms of the possibilities of tasks.

*2.14 Abstract constructors*

The classic philosophical puzzle of the 'ship of Theseus' concerns the legend that after Theseus's adventure in Crete, the Athenians made his ship a monument, replacing any plank that decayed. Eventually, none of the original planks were left. Was it still the same ship, and if not, when did it stop being so?

Though the original puzzle has little substance (being hardly more than an essentialist exercise about the meaning of 'same'), it does illustrate a fact about the physical world which, in the prevailing conception, is counter-intuitive. Let us approximate the ship as a substrate on which a constructor, the city of Athens, is repeatedly performing the task of restoring it to its original state. To prevent it from gradually changing shape with successive renovations, some blueprint must exist, specifying the correct shape. So Athens must be a *programmable* constructor, with the blueprint in a program whose physical representation (say, on a scroll) is an additional substrate that Athens must keep from deteriorating. Another part of the program would be the *ideas* that cause generations of Athenians to keep doing that. And since different Athenians perform the task in each generation, they, too, are among the substrates on which the task is being repeatedly performed.

So, since the ship, the scroll, and Athens are all substrates, what is the constructor? By hypothesis, none of the physical objects effecting the repair survives unchanged for many generations. Only the program does: the abstract information expressed in the scroll and the ideas. So that information meets the definition of a constructor more closely than anything else in the situation. It is an abstract constructor.

There seem to be constructor-theoretic laws, such as the interoperability principle for information (Section 2.6 above), that refer directly to abstract constructors. If there are such laws, then abstract constructors cannot be omitted from fundamental physics in the constructor-theoretic conception. Readers who baulk at the idea of an abstraction *causing* something physical need only substitute a weaker term for 'can cause' in the definition of a constructor in Section 1.1, such as 'can cause, or codes for'.

*2.15 Knowledge*

The most important kind of abstract constructor is *knowledge*. Knowledge is information which, once it is physically instantiated in a suitable environment, tends to cause itself to remain so: it survives criticism, testing, random noise, and error-correction. (Here I am adopting Popper's (1972) conception of knowledge, in which there need be no knowing subject.) For example, the knowledge encoded in an organism's DNA consists of abstract genes that cause the environment to transform raw materials into another instance of the organism, and thereby to keep those abstract genes, and not mutations or other variants of them, physically instantiated, despite the mutation and natural selection that keep happening. Similarly, the ideas constituting the abstract constructor for preserving the ship of Theseus would have had to include not only some relatively arbitrary information about the historical shape of the ship, but also knowledge of how to cause Athenians to preserve those ideas themselves through the generations, and to reject rival ideas.

Now consider again the set of all physically possible transformations. For almost every such transformation, the story of how it could happen is the story of how knowledge might be created and applied to cause it. Part of that story is, in almost all cases, the story of how *people* (intelligent beings) would create that knowledge, and of why they would retain the proposal to apply it in that way while rejecting or amending rival proposals (so a significant determinant is *moral* knowledge). Hence, from the constructor-theoretic perspective, physics is almost entirely the theory of the effects that knowledge (abstract constructors) can have on the physical world, via people. But again, the prevailing conception conceals this.

*2.16 Emergent and scale-independent laws*

Almost all physical processes involving large numbers of elementary particles are intractably complex. But for a tiny but important minority, that complexity resolves itself into simplicity at a higher level – a phenomenon known as *emergence*. Specifically, certain sets of collective phenomena can be explained in terms of emergent laws relating them only to each other, without reference to the underlying particles and laws.

Some forms of emergence are unproblematic under the prevailing conception because their laws follow logically from low-level laws, either as well-defined approximations (such as the gas laws) or exactly (such as the law of motion of a centre of mass). But in some cases an exact emergent law appears to exist, yet not to follow from lower-level laws. The principle of testability and the second law of thermodynamics are examples. These do conflict with the prevailing conception. It could be that they are exactly true at all scales but that, as with the laws of chemistry, that is only manifest if they are expressed in constructor-theoretic terms – the former as a principle of constructor theory and the latter within a subsidiary theory, thermodynamics, in a constructor-theoretic formulation.

In this paper I use the term 'laws of nature' instead of 'laws of physics' where I want to stress the inclusion of emergent laws. If the constructor-theoretic approach were to result in scale-independent emergent laws, the two concepts would be unified. However, not all sciences dealing in emergent phenomena seek *laws* of nature. For example, historians trying to understand (say) the effect of Gutenberg's invention on the Enlightenment will not necessarily ever do so by deducing it from the dichotomy between possible and impossible historical transformations – just as today they do not necessarily seek a law of motion for intellectual movements. So even complete success in discovering scale-invariant *laws* would not in principle reduce all sciences to constructor-theoretic physics.

### 2.17 Complexity theory

The theory of computation draws a qualitative distinction between computable and non-computable, but it has an additional, quantitative branch, complexity theory, which is about the resources required to perform computable tasks. Its usefulness is limited by the fact that it cannot address physical resources (such as mass and time) but only abstract proxies (such as memory capacity and number of computational steps), and then only asymptotically in the 'size' of the input, and only up to an unknown polynomial function.

So there is scope for a 'construction complexity theory' that generalises computational complexity theory to cover all construction tasks, and is in principle capable of addressing physical resource requirements accurately. Remarkably, that theory is none other than constructor theory itself. For example, consider the task of transmuting a mass *m* of hydrogen into at least a mass *M* of gold, using energy at most *E*. Relativity tells us that this task is impossible when $E < (M-m)c^2$, but presumably no constructor can closely approach that limit. A near-optimal constructor for this task might work something like this: when $E \gg mc^2$ it uses the excess energy to power a particle accelerator that converts the hydrogen into a slightly smaller mass of gold. Then it converts the remaining energy into particle-antiparticle pairs, discards the antiparticles, and converts the particles into gold using the accelerator again. When $E \ll mc^2$, it augments the energy *E* with energy extracted from the hydrogen, say using a nuclear fusion reactor, which it again uses to power an accelerator to transmute the fusion waste products into gold. The diagram shows, for fixed *M*, the region of the *m–E* plane for which the task is possible (schematically: the real diagram may have intricate detail).

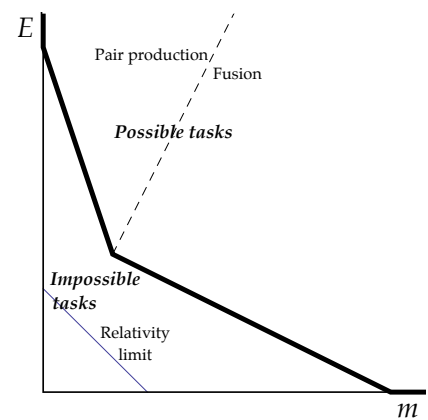

Now, this diagram, in which the heavy line partitions a set of tasks into the possible and the impossible, can also be regarded as expressing a quantitative 'construction-complexity' result: the same line is also a *graph* of the energy required to transmute a mass *m* of hydrogen into a fixed mass of gold. This natural unification of complexity with computability when they are both generalised suggests that a deeper theory exists along the lines that I am advocating.

*2.18 Pragmatic considerations*

Disputes about whether certain simply-defined construction tasks are possible or impossible are common. Examples are: building a universal quantum computer (Landauer 1995); achieving controlled fusion; immortality (de Grey 2007); 'molecular assemblers' (Drexler 1995); and the issue of how long Moore's law can continue to hold. The possibility of nuclear power and of aeroplanes were once similarly controversial. To date, such disputes have never been grounded in fundamental theory; they are more about the participants' intuitions than science. But a valid argument that a particular task is impossible must ultimately rely on some law of nature; and any argument that a particular task is possible when no constructor for it is yet known must be appealing to some constructor-theoretic principle, at least implicitly, for while the prevailing conception can rule out tasks (given existing laws), it cannot rule them in until after a constructor for them has been designed.

## 3. Development of the theory

### 3.1 Pharaonic tasks

Imagine a task $\mathcal{A}$ for which the laws of nature would permit (good approximations to) a constructor $\mathbf{C}_\mathcal{A}$ to continue to exist if one already existed, but would not allow one to be constructed from naturally-occurring resources. That would imply $\mathcal{A}^\times$ no less than if there were a limit on how accurately the transformations specified by $\mathcal{A}$ could be achieved. Yet, what I have called 'performing' $\mathcal{A}$ does not include constructing $\mathbf{C}_\mathcal{A}$, nor obtaining the resources required to construct it. Let me call the task that includes all those preliminaries the *pharaonic*[1] version of $\mathcal{A}$. Performing it requires other constructors, whose construction must in turn be possible (in the sense defined in Section 1.1). To avoid an infinite regress, $\mathcal{A}^\checkmark$ must imply that every chain of transformations contributing to the construction of $\mathbf{C}_\mathcal{A}$ begins with naturally-occurring constructors and substrates.

What resources are 'naturally occurring' depends partly on local conditions, so the possibility of a pharaonic task does too. Yet for ordinary tasks it must depend only on laws of nature. For example, suppose that causing a particular transformation requires conditions that existed in the early universe but could not be provided today regardless of what knowledge was brought to bear. Then whether the transformation could take place would be a function of cosmological epoch. The principles of constructor theory imply that any such dependence must be entirely implicit – i.e. due to physical variables, such as pressure or spacetime curvature, of local systems, and not to time as such. Those systems would be substrates of the relevant tasks, and the possibilities of those tasks would be timeless truths, as required.

---

[1] Referring to the Biblical story of the pharaoh who ordered that the Israelite slaves find their own brick-making materials instead of transforming a given supply.

Now, the idea of constructor theory is that the set of all truths about which tasks are possible and impossible, and why, encompasses the whole of physics, including all facts about which transformations could take place during each cosmological epoch. In other words, the theory of pharaonic tasks must in principle follow entirely from that of ordinary tasks. How can it?

Note that the distinction between ordinary and pharaonic tasks corresponds roughly to that in the prevailing conception between laws of motion and the initial state at the Big Bang. Laws of motion do constrain the initial state (for example, if they are differential equations, the initial data must be sufficiently differentiable), but not enough to determine it. But principles of constructor theory requiring certain tasks to be possible (such as lengthy computations, or tests of laws of nature) impose draconian constraints on the initial state. So it may well be that, together with the subsidiary theories, they determine the possibilities of all relevant pharaonic tasks, and thence also the initial state as an emergent consequence, as suggested in 2.5 above.

*3.2 The possible-impossible dichotomy*

Impossible tasks cannot be effected regardless of what knowledge is brought to bear. But if the right knowledge is applied, any possible task $\mathcal{A}$ can necessarily be effected in practice with arbitrary accuracy short of perfection, for the following reason: If, in the event, nothing will ever perform $\mathcal{A}$ with better than a particular accuracy, that may just be an accidental regularity and so would not contradict $\mathcal{A}^{\checkmark}$. But suppose that there is some obstacle to doing so. If the obstacle is parochial – i.e. we do not have enough knowledge or wealth (see 3.9 below) – then, again, that does not contradict $\mathfrak{A}'$. But if it were an obstacle that *must* remain insuperable regardless of what knowledge may be brought to bear, then a statement that that was so would express a law of nature. And that would contradict $\mathfrak{A}'$. So there cannot be such an obstacle. That is the motivation for calling anything that is not forbidden by laws of nature 'possible'.

*3.3 Impossible states*

Similarly, impossible *substrates*, or impossible states of substrates, are those that are forbidden by laws of nature. The task of transforming a possible state into an impossible one is of course impossible, but impossible states may nevertheless appear in the formalism of subsidiary theories, and transforming one impossible state into another may well be a possible task. For example, Maxwell's electrodynamics denies the existence of magnetic monopoles, but also predicts how, if they existed, they would interact with electromagnetic fields. This allows the construction of monopole-detecting instruments, which are constructors capable of testing the prediction that no monopoles exist.

*3.4 Towards a purely task-based constructor theory*

Consider a subsidiary theory about a certain substrate. Among the facts that this theory presents to constructor theory might be a set of tasks $\mathfrak{T}_R$ parameterised by the set of all $3 \times 3$ matrices $R$, with the property that for any such matrices $R_1$ and $R_2$, $\mathfrak{T}_{R_1} \mathfrak{T}_{R_2} = \mathfrak{T}_{R_1 R_2}$, and that $\mathfrak{T}_{\tilde{R}} = \mathfrak{T}_{R^{-1}}$. From this constructor-theoretic information, we could deduce that these tasks represent some sort of linear transformation of physical parameters of the substrate. In addition, the theory might say that $\mathfrak{T}_R^{\checkmark}$ whenever $R$ is an orthogonal matrix, and $\mathfrak{T}_R^{\mathbf{x}}$ when it is not – in which case we could begin to guess that the tasks constitute rotations in some three-dimensional space. In this way, the algebra of tasks and their partition into possible and impossible would implicitly include information about properties of the substrate's states.

Similarly, they could encode information about the laws governing the substrates. For example, suppose that the substrates are subject to an additive conservation law. That law will prohibit a certain class of tasks. Let $\{x \to y\}^{\mathbf{x}}$ be such a prohibition. Then we must also have $\{y \to x\}^{\mathbf{x}}$ because that would violate the same law in the opposite direction. However, for some pairs of tasks $\mathfrak{A}$ and $\mathfrak{B}$ that both violate the same conservation law by the same amount, it could be that $(\mathfrak{A}^{\sim} \otimes \mathfrak{B})^{\checkmark}$ and $(\mathfrak{A} \otimes \mathfrak{B}^{\sim})^{\checkmark}$ – indeed, if that is not so, it is because some other law forbids those tasks.

Different patterns of possible and impossible tasks result from different types of laws. For instance, the second law of thermodynamics must be reflected in patterns like $\mathfrak{A}^{\checkmark} \& \mathfrak{A}^{\sim \times}$.

In a task-based formalism, subsidiary theories would also be required to define a relation of union between tasks (corresponding to set-theoretic union in the notation of this paper). Then states would correspond to tasks $\mathfrak{A}$ that equal their transposes and have no proper *subtasks*, so they would have the form $\{a \to a\}$ in the notation of this paper.

*3.5 Distinguishing law from accidental regularity*

There is a longstanding philosophical controversy about how, or whether, laws of nature differ from mere summaries of factual regularities (Swartz 1995). Constructor theory's fundamental distinction between impossible and possible is between tasks that are forbidden by laws of nature and tasks that are not, even if they are never performed. So constructor theory is incompatible with the position that laws of nature are just summaries of what happens.

*3.6 A central role for the impossible*

Much of physics in the constructor-theoretic conception is about impossible tasks, and tasks on impossible states (Section 3.3). This central role for the impossible is not only a formal implementation of the Popperian idea that the content of a scientific theory is in what it forbids. It is also an important difference between the constructor-theoretic conception of the physical world and the prevailing one: what *actually* happens is seen as an emergent consequence of what *could* happen, rather than vice-versa.

*3.7 Repertoires of programmable constructors*

In order to evaluate a computable function of a particular input, a computer needs enough working memory to store the intermediate results. Since, in reversible computations, those results are erased by the time the output is delivered, that memory is not depleted by the computation and can be re-used, so it is both a substrate and a constructor. We may call non-programmable constructors that are used as substrates of other constructors *ancillas*, generalising the existing use of the term in the theory of computation.

For many programmable computers, including every universal one, the memory requirement of computations in the computer's repertoire has no upper bound. Therefore such a computer cannot be defined as including all its ancillas, and the repertoire of a programmable computer must be defined as the set of computations it can be programmed to perform *if* it is given an unlimited supply of additional memory.

Another reason for not counting ancillas among the resource requirements of computations is that if one does count them, there are no universal computers. Let **C** be a computer capable of performing a computational task

$$\mathfrak{A} = \bigcup_{i \in \mathbb{Z}} \{i \to f(i)\}, \tag{11}$$

using $b(i)$ memory-ancillas of a given capacity and in $t(i)$ steps when the input is $i$. Then a universal computer programmed to simulate **C** may require up to $p(t(i))$ steps and up to $q(b(i))$ ancillas, where $p$ and $q$ are polynomial functions, *every time* it performs $\mathfrak{A}$ with input $i$. Since no computer architecture optimises the memory and time requirements of every computational task, there would be no universal computer if limits on time and memory, and hence on ancillas, were included in the definitions of computational tasks.

In constructor theory a stricter conception of universality is possible, because when a programmable *constructor* is programmed to mimic another constructor **C**, it may begin by constructing an instance of **C**, to which it directs subsequent inputs, so that from then on it performs **C**'s task using much the same resources that **C** would.

Hence the repertoire of a programmable constructor **P** can be defined as the set of tasks that **P** can be programmed to be capable of performing, *up to a constant amount of naturally-occurring resources*. More precisely, the overhead of programming **P** to be capable of performing $\mathfrak{A}$ is a constant $c(\mathbf{P},\mathfrak{A})$, independent of how often $\mathbf{P}$ will then be called upon to perform $\mathfrak{A}$, and which inputs for $\mathfrak{A}$ it is given.

### 3.8 Universal constructors

A *universal constructor* is a programmable constructor whose repertoire is the union of all repertoires of all possible programmable constructors. It seems plausible that a universal constructor can exist. Consider, first, any constructor that human engineers could be instructed how to build. Those instructions could presumably be translated into a program for programmable robots with humanoid manipulative abilities. And since any such robot $\mathbf{R}$ could be programmed to build any other, it would be a constructor whose repertoire included all human-achievable tasks. If $\mathbf{R}$ were not a universal constructor, there would have to be a task $\mathfrak{A}$ that is not human-achievable but nevertheless possible. Given the necessity for knowledge creation, performing $\mathfrak{A}$ would presumably require knowledge that could be created by some naturally occurring non-human entities. But by the foregoing argument, $\mathfrak{A}$ would also be in the repertoire of some programmable robot $\mathbf{R}'$, buildable by those entities. The robots $\mathbf{R}$ and $\mathbf{R}'$, plus a device for presenting any input to the appropriate one and programming it with the relevant program, would together constitute a programmable constructor whose repertoire contained both repertoires. It seems implausible that there could be laws of nature forbidding those entities and us from cooperating in that way.

Plausibility arguments aside, I conjecture that the existence of a universal constructor can be proved from independently-motivated principles of constructor theory.

A universal constructor will typically perform a task by first constructing a more specialised constructor. That constructor may do the same. Define the *n*'th-level repertoire of a constructor as the set of tasks it can perform without causing more than *n* levels of constructor to be built. For each universal constructor, this induces a classification of all possible tasks, according to which level of the constructor's repertoire it is in. Presumably all universal constructors reach universality at a finite level; I conjecture that there is a universal upper bound on that level.

A further unification in constructor theory is that unlike with computers, there is no intermediate universality at a classical (non-quantum) level. If constructing a universal quantum constructor is possible, the classical robots envisaged above can presumably be programmed to construct it.

*3.9 Wealth*

In constructor theory it is natural to define the *wealth* of an entity in a non-anthropocentric way as the set of transformations that the entity would be capable of performing without generating new knowledge. Wealth has always consisted fundamentally of knowledge, even though it has been limited by the capacity of relatively fixed installations for harnessing naturally occurring resources. Once universal constructors exist, it will consist almost entirely of knowledge.

*3.10 Locality*

The requirement (Section 1.1) that a construction task specify only intrinsic properties of the substrate has an important consequence that is already known in existing physics, namely Einstein's *principle of locality* (1949): for any two spatially separated physical systems $S_1$ and $S_2$, 'the real factual situation of the system $S_2$ is independent of what is done with the system $S_1$'. This is a necessary condition for tasks to be composable into networks as envisaged in Section 1.2. Expressed as a principle of constructor theory, it is: for any two substrates $S_1$ and $S_2$, there is a way of describing their states such that if $S_1$ is in state $x$ and $S_2$ is in state $y$, the state of $S_1 \oplus S_2$ is the ordered pair $(x,y)$. If tasks, not states, turn out to be the primitive entities of physics, as suggested in Section 1.1, then the mere principle that all networks of tasks are themselves tasks has the same effect.

The principle of locality holds unproblematically in classical physics but quantum entanglement is often claimed to violate it. However, there exist formally local formulations of quantum theory (Deutsch & Hayden 2000; Deutsch 2012), which are compatible with constructor theory.

When subsidiary theories define locality, they automatically define something about simultaneity too: two substrates can be consistently regarded as being *simultaneously* in states $x$ and $y$ respectively if and only if the composite system of both of them is in *some* state – some 'real, factual situation' – $(x, y)$. For instance, according to relativity, 'simultaneous' in this sense means 'separated by a spacelike interval'; according to quantum theory, it means that any two (Heisenberg-picture) observables, one from each substrate, commute (Deutsch & Hayden 2000).

*3.11 Interoperability principles*

In 2.8 I mentioned two branches of physics – geometry and computation – whose fundamental laws were once mistaken for self-evident truths. That happened because the true laws are, in both cases, nearly substrate-independent. In Euclid's geometry, the sum of the angles of a triangle is 180° whether the vertices are marked by candles or lasers. In Euclid's highest-common-factor algorithm, one obtains the same sequence of remainders whether one executes it with beads or electrons. When a class of substitutions of substrates leaves the possibilities of all the tasks in a certain class unchanged, a statement of that substrate-independence is an *interoperability law* or principle. In 2.6 I discussed the interoperability principle for information. My speculation about thermodynamics in 2.3 suggests that there are interoperability principles for work and heat.

Just as knowledge is an abstract constructor (2.14), so every interoperability law is associated with some abstract constructor or abstract substrate, or both. For example, if a task requires work to be converted into heat, the work is an abstract substrate; otherwise, it is an abstract constructor – and indeed, chemists do sometimes call heat a 'catalyst' for exothermic reactions.

*3.12 Time*

In both quantum theory and general relativity, time is treated anomalously. The problem in both theories is that time is not among the entities to which the theory attributes objective existence (namely quantum observables and geometrical objects respectively), yet those entities change with time. So there is widespread agreement that there must be a way of treating time 'intrinsically' (i.e. as emerging from the relationships between physical objects such as clocks) rather than 'extrinsically' (as an unphysical parameter on which physical quantities somehow depend). But this is difficult to accommodate in the prevailing conception, every part of which (initial state; laws of motion; time-evolution) assumes that extrinsic status. Attempts have been made to reformulate both theories with intrinsic time (Page & Wootters 1983; Barbour 1999, 2012), but have not yet achieved much generality. In the constructor-theoretic conception, it is both natural and unavoidable to treat both time and space intrinsically: they do not appear in the foundations of the theory but are emergent properties of classes of tasks whose substrates include 'rods' and 'timers'. (That constructor theory allows for serial composition to be noncommutative, and for a task and its transpose to be different, may be considered a built-in *direction* of time, but not a time parameter.)

Extrinsically, a timer might be defined as a substrate $\mathbf{T}_t$ which, if left isolated in a suitable initial state, emits some information when a time $t$ has passed. But that condition about time 'passing' is not constructor-theoretic. Some constructor-theoretic content can be added by noting that there is an interoperability law for timers: any timer $\mathbf{T}_t$ is guaranteed to be replaceable, in any construction, by any other timer $\mathbf{T}'_t$ with the same parameter $t$, without changing the task that the constructor performs. A fully intrinsic constructor-theoretic conception of timers would be in terms of tasks that take time. For example, let $\mathbf{R}$ be a non-conducting rod of length $l$ with an additional electron. Let $x$ and $y$ be states in which the electron is located at either end of $\mathbf{R}$, and let $\mathbf{C}_1$ and $\mathbf{C}_2$ be constructors for the task $\mathcal{T}_R = \{x \to y, y \to x\}$. Associate them in parallel, with an additional apparatus ensuring that when $\mathbf{C}_1$ halts, it incapacitates $\mathbf{C}_2$. Define 'faster' such that if that composite constructor is capable of performing the task $\mathcal{T}_R \otimes \mathcal{T}_R$, but the same does not hold if $\mathbf{C}_1$ and $\mathbf{C}_2$ are interchanged, then $\mathbf{C}_2$ is faster than $\mathbf{C}_1$. Then, if $\mathbf{C}$ is a constructor for $\mathcal{T}_R$ such that no other constructor for $\mathcal{T}_R$ is faster than it, it can act as a timer. For instance, if the subsidiary theory was relativity, $\mathbf{C}$ would time a period $t = l/c$ where $c$ is the speed of light.

*3.13 Halting and synchronisation*

For a computer to perform a computational task, it must not only place the right information into some memory register, it must give some positive indication that it has done so. Otherwise, a program that ignored its input and simply counted indefinitely would qualify as computing any integer-valued function of the input, computable or not. The same holds for constructions: a constructor must give such an indication, otherwise the user does not have the information necessary to avoid taking the cake out of the oven before it is cooked (i.e. the construction is still under way), or when it is burnt or stale (i.e. the substrate has changed after being in its correct output state). Giving this indication is usually called 'halting', but in the case of networks of constructions (not tasks) it is more appropriately called *synchronisation*.

The inputs of a construction must be presented to the constructor simultaneously (in the sense of 3.9 above) and the constructor must deliver its outputs simultaneously. For instance, a pair of constructors for $\mathfrak{A}$ and $\mathfrak{B}$ respectively will not necessarily perform $\mathfrak{A} \otimes \mathfrak{B}$ if they are run in parallel, because if the difference in their running times is too great, they may not produce the combination of the substrates in a legitimate output state. Similarly, the task of presenting a constructor with its input, or of conveying an intermediate substrate from one constructor in a network (*of constructors*) to another is not necessarily trivial. But if the composition principle holds in the form I suggested (Section 1.2), then although composing particular constructors may be impossible, this never makes a regular network of possible *tasks* impossible: either the net effect can be achieved in a different way, or the synchronisation can be effected by means such as moving substrates rapidly and making use of time dilation.

*3.14 Imperfect constructors*

As I remarked in 1.1, no perfect constructor can exist in nature. If nothing else, thermal noise causes random transitions, so there is always a non-zero probability that the wrong transformation will happen or that the device will undergo a change large enough to destroy its functionality, thus disqualifying it from being a constructor. So that raises the question: constructor theory expresses *laws* of physics in terms of the possibilities of tasks, but how could one describe actual physical processes (i.e. imperfect constructions) in constructor-theoretic terms?

For example, consider the task

$$\mathfrak{A} = \bigcup_{\substack{h \in M(\text{'heads'}) \\ t \in M(\text{'tails'})}} \{h \to t\} \tag{12}$$

performed on a coin, where $M(\text{'heads'})$ is the set of all states in which we deem it to be at rest in the 'heads' orientation, and similarly for $M(\text{'tails'})$. Assume for the sake of argument that $\mathfrak{A}^{\checkmark}$, since no conservation law or law of motion forbids $\mathfrak{A}$, and $\mathfrak{A}$ is logically reversible. Turning the coin in real life will differ from a performance of $\mathfrak{A}$ in two basic ways: the coin and the approximate constructor will differ from their idealised specifications both before and after the process; and additional substrates will interact with them during it.

For instance, turning the coin involves adding and removing kinetic energy. In the limit of performing $\mathfrak{A}$, all this energy would come from the constructor and then be returned to it. But in any realisable situation some energy would be dissipated, thus necessitating an energy source and sink, which constitute one or more additional substrates. Also, thermal noise will be present, so the (mixed) quantum state of the coin will include, with non-zero coefficients, some states not in $M(\text{'heads'})$ before the process, and states not in $M(\text{'tails'})$ afterwards. However, there exists a class of more lenient tasks $\mathfrak{A}_\varepsilon$, where the parameter $\varepsilon$ expresses some upper limit on the amount of energy that may be dissipated in performing $\mathfrak{A}_\varepsilon$, and on the amount by which the state of the coin may differ from legitimate input and output states of $\mathfrak{A}$. $\mathfrak{A}_\varepsilon$ would also have to refer to the states of further substrates in the coin's environment, such as the table on which it may be resting. Performing $\mathfrak{A}_0$ preserves the energy and all other relevant variables of all substrates other than the coin, and hence those substrates can then be considered part of the constructor. Hence we should identify $\mathfrak{A}_0 = \mathfrak{A}$ in constructor algebra even though $\mathfrak{A}_0$ has different substrates from $\mathfrak{A}$.

Suppose that $\mathfrak{A}_\varepsilon$ allows completely unconstrained outputs in a small proportion of universes. For example, its legitimate output states might be all quantum states of the substrates such that $\langle \hat{P}_{\text{'tails'}} \rangle \geq 1 - \varepsilon$, where $\hat{P}_{\text{'tails'}}$ is the projector for all states in $M(\text{'tails'})$. Then the effect *on the substrates* of performing $\mathfrak{A}_\varepsilon$ *perfectly* can be achieved in reality. But the entity causing that effect will still deteriorate, so it is still not a constructor for $\mathfrak{A}_\varepsilon$. To describe the imperfect constructors themselves, one can consider them as substrates **C** of tasks of the form 'cause **C** to perform $\mathfrak{A}_\varepsilon$ and to remain capable of doing so again'.

*3.15 Are we universal constructors?*

I guess that neither a typical human nor human civilisation as a whole approximates a universal constructor – not because we are something less but because, I hope, we are something more: we cannot be programmed – and especially not programed to carry out arbitrary instructions for an arbitrarily long time – because we may not want to.

Since we do not have hardware to upload arbitrary information into our brains as programs, the issue comes down to this: for each possible task $\mathfrak{A}$, does there exist *information* $M(\mathfrak{A})$ such that if we received a message instantiating $M(\mathfrak{A})$, from space, followed by the substrates of $\mathfrak{A}$ in a legitimate input state, and a supply of naturally-occurring raw materials, we would reliably transform the substrates to the corresponding output states, and then be ready to do so again for another possible task? What would $M(\mathfrak{A})$ have to say, to cause us to do this?

Presumably the effective part of $M(\mathfrak{A})$ would be in its preamble. Could it be a threat: 'we are immensely powerful and will destroy you unless you obey the following instructions faithfully…'? Or a trade: 'we offer you the secret of immortality if you pass this test…'? Presumably neither would be sufficient to cause us to be just as capable of performing $\mathfrak{A}$ again afterwards in response to another instance of $M(\mathfrak{A})$. Among other things, $M(\mathfrak{A})$ would have to cause changes in our civilisation that prevented our ideas from moving in directions that would make us disinclined to obey future commands. Perhaps there exists some way of fooling us into making something that would destroy us after causing us to build a more straightforward universal constructor. But again, could that be done with high reliability? I think that in reality, our creativity makes it implausible that we approximate a universal constructor very closely, despite our presumed ability to build one.

If a universal constructor is possible, there must be a smallest one. It would be interesting to know what it is, and how it works. Is it at the scale of molecules? If so, it may become the centrepiece of nanotechnology. If it is much larger, then general-purpose nanotechnology will never be independent of macroscopic control and support systems. Will it just be a curiosity (like the simplest Turing machine)? Or will it rapidly become the commonest pattern of matter in the universe, the vehicle by which knowledge comes to dominate everything that happens?

## 4. Conclusion

The principles of constructor theory that I have proposed may be false. For instance, the composition principle, in the form stated in Section 1.2, may only be an approximation. But if the idea as a whole is false, something else will have to remedy the deficiencies of the prevailing conception. Something else will unify emergent-level laws such as the Turing principle and the principle of testability with the other laws of nature. Something else will provide an exact statement of the second law of thermodynamics, and a full statement of the content of conservation laws. A different approach will generalise the theory of computation and von Neumann's constructor theory, and support laws about substrate-independent quantities such as information. And incorporate into fundamental physics the fact that the most significant quantity affecting whether physical transformations happen or not is knowledge.

**Acknowledgements**

I am grateful to Simon Benjamin for pointing out some subtleties including the non-existence of a universal classical constructor, to him and Mark Probst for illuminating conversations on the themes of this paper, to Alan Forrester and two anonymous referees for numerous useful suggestions, and especially to Chiara Marletto for incisive criticism of earlier versions of the theory and of earlier drafts of this paper.

**References**
Barbour, J. 1999 *The End of Time*, Weidenfeld and Nicolson.
Barbour, J. 2012 *Shape Dynamics. An Introduction* in *Quantum Field Theory and Gravity*, Finster, F. et al. (eds) Birkhäuser, Basel.
Brown, H.R. & Timpson, C.G., 2006 in *Physical Theory and Its Interpretation: Essays in Honor of Jeffrey Bub*, W. Demopoulos & I. Pitowsky (eds.), Springer; pp. 29-41.
de Grey, A. 2007 in Rae, M. (ed.) *Ending Aging* St. Martin's Press, New York.
Deutsch, D. 1985 *Quantum theory, the Church-Turing principle and the universal quantum computer* Proc. R. Soc. **A400** *A* **400** 97-117

Deutsch, D. 1997 *The Fabric of Reality* Allen Lane The Penguin Press, London.
Deutsch, D. 2002 *The structure of the multiverse* Proc. R. Soc. **A458**, 2911-23.
Deutsch, D. 2011 *The Beginning of Infinity.* Allen Lane, The Penguin Press, London.
Deutsch, D. 2012 *Vindication of quantum locality* Proc. R. Soc. **A468** 531-44.
Deutsch, D. & Hayden, P. 2000 *Information flow in entangled quantum systems* Proc. R. Soc. **A456** 1759-74.
Drexler, K.E. 1995 *Molecular manufacturing: perspectives on the ultimate limits of fabrication* Phil. Trans. R. Soc. **A353** 323-31.
Einstein, A. 1908 Letter to Arnold Sommerfeld, Document 73 in The Collected Papers of Albert Einstein, Vol. 5, The Swiss Years: Correspondence, 1902–1914 (English Translation Supplement, M. J. Klein, A. J. Kox and R. Schulmann (eds.), Princeton University Press, Princeton. Translated by A. Beck., 1995.
Einstein, A. 1949 quoted in *Albert Einstein: Philosopher, Scientist*, P.A. Schilpp, Ed., Library of Living Philosophers, Evanston, 3rd edition (1970), p85.
Hume, D 1739 *A Treatise of Human Nature* Clarendon Press, Oxford (2007).
Kant, I. 1781 *Transcendental Exposition of the Concept of Space* in *Critique of Pure Reason*.
Landauer, R. 1995 *Is Quantum Mechanics Useful?* Phil. Trans. R. Soc. **A353** 367-76.
Page, D.N. & Wootters, W. 1983 *Evolution without evolution: Dynamics described by stationary observables* Phys. Rev. **D27** 12, 2885-892.
Popper, K.R. 1959 *The Logic of Scientific Discovery*, Routledge.
Popper, K.R. 1963 *Conjectures and refutations*, Routledge.
Popper, K.R. 1972 *Epistemology Without a Knowing Subject* in *Objective Knowledge: An Evolutionary Approach* ch. 3, Oxford University Press.
Putnam, H. 1974, *The 'Corroboration' of Theories* in Schilpp, P.A. 1974, *The Philosophy of Karl Popper* **1** 221, Open Court, La Salle, Illinois. See also Popper's reply *Putnam on "Auxiliary Sentences", Called by Me "Initial Conditions" loc. cit.* **2** 993.
Quine, W.V.O. 1960, *Word and Object* p189 MIT Press.
Russell, B. 1913, *On the Notion of Cause* Proc. Aristotelian Soc. New Series **13** (1912 - 1913) 1-26.
Swartz, N. 1995, *A Neo-Humean Perspective: Laws as Regularities*, Cambridge University Press.
Turing, A.M. 1936, *On Computable Numbers, with an Application to the Entscheidungsproblem* Proc. Lond. Math. Soc. 2 **42** (1) 230–65.
von Neumann, J., 1948 *The General and Logical Theory of Automata*, Hixon Symposium, September 20, 1948, Pasadena, California